\documentclass[twocolumn,aps,prb,superscriptaddress,citeautoscript]{revtex4}

\usepackage{graphicx}
\usepackage{dcolumn}
\usepackage{bm}
\usepackage{amsmath}
\usepackage[T1]{fontenc}
\usepackage{sidecap}
\usepackage{float}
\usepackage{caption}
\usepackage{booktabs}
\usepackage{hyperref}
\hypersetup{
    colorlinks,
    linkcolor={black},
    citecolor={blue},
    urlcolor={blue}
}

\begin{document}

\preprint{}

\title{Spin-orbit torque induced magnetisation dynamics and switching in CoFeB/Ta/CoFeB system with mixed magnetic anisotropy}

\author{Stanis\l{}aw \L{}azarski}
 \email{lazarski@agh.edu.pl}
\affiliation{AGH University of Science and Technology, Department of Electronics, Al. Mickiewicza 30, 30-059 Krak\'{o}w, Poland}
\author{Witold Skowro\'{n}ski}
 \email{skowron@agh.edu.pl}
\affiliation{AGH University of Science and Technology, Department of Electronics, Al. Mickiewicza 30, 30-059 Krak\'{o}w, Poland}
\author{Krzysztof Grochot}
\affiliation{AGH University of Science and Technology, Department of Electronics, Al. Mickiewicza 30, 30-059 Krak\'{o}w, Poland}
\affiliation{AGH University of Science and Technology, Faculty of Physics and Applied Computer Science, Al. Mickiewicza 30, 30-059 Krak\'{o}w, Poland}
\author{Wies\l{}aw Powro\'{z}nik}
\affiliation{AGH University of Science and Technology, Department of Electronics, Al. Mickiewicza 30, 30-059 Krak\'{o}w, Poland}
\author{Jaros\l{}aw Kanak}
\affiliation{AGH University of Science and Technology, Department of Electronics, Al. Mickiewicza 30, 30-059 Krak\'{o}w, Poland}
\author{Marek Schmidt}
\affiliation{Institute of Molecular Physics, Polish Academy of Sciences, ul. Smoluchowskiego 17, 60-179 Pozna\'{n}, Poland}
\author{Tomasz Stobiecki}
\affiliation{AGH University of Science and Technology, Department of Electronics, Al. Mickiewicza 30, 30-059 Krak\'{o}w, Poland}
\affiliation{AGH University of Science and Technology, Faculty of Physics and Applied Computer Science, Al. Mickiewicza 30, 30-059 Krak\'{o}w, Poland}

\date{\today}

\begin{abstract}
Spin-orbit torque (SOT) induced magnetisation switching in CoFeB/Ta/CoFeB trilayer with two CoFeB layers exhibiting in-plane magnetic anisotropy (IPMA) and perpendicular magnetic anisotropy (PMA) is investigated. Interlayer exchange coupling (IEC), measured using ferromagnetic resonance technique is modified by varying thickness of Ta spacer. The evolution of the IEC leads to different orientation of the magnetic anisotropy axes of two CoFeB layers: for thicker Ta layer where magnetisation prefers antiferromagnetic ordering and for thinner Ta layer where ferromagnetic coupling exists. Magnetisation state of the CoFeB layer exhibiting PMA is controlled by the spin-polarized current originating from SOT in $\mu m$ sized Hall bars. The evolution of the critical SOT current density with Ta thickness is presented, showing an increase with decreasing $t_\mathrm{Ta}$, which coincides with the coercive field dependence. In a narrow range of $t_\mathrm{Ta}$ corresponding to the ferromagnetic IEC, the field-free SOT-induced switching is achieved.
\end{abstract}

\maketitle

\section{Introduction}
Conventional electronic devices based only on the electric charge are facing energy efficiency issues due to the use of high current densities needed to control the device state that dissipates a lot of heat. Moreover, digital data stored in the form of electric charge for example in dynamic and static random access memories are volatile and  need periodic refreshment. Spintronic devices use both charge and spin of the electron to process and store information. Magnetism-based spintronic elements are by nature nonvolatile and potentially more energy-efficient, nonetheless, there is still a need for a further development of magnetisation control mechanisms.\cite{ikegawa_2020} Digital information can be encoded in the magnetisation state of nanoscale magnet in a non-volatile way. However, research on various control mechanisms have recently received extensive attention due to the demand for scalable and low-energy write and read mechanisms.\cite{bhatti_2017} 
Outstanding developments has been made in switching magnetic tunnel junctions (MTJs) and spin-valve structures via spin-transfer torque (STT)\citep{slonczewski_1996,berger_1996}, electric fields\cite{ohno_2000,kanai_2012} and recently vastly explored spin-orbit torque (SOT).\cite{miron_2011,liu_2012PRL,emori_2013} 
MgO-based MTJs with in-plane magnetic anisotropy have been initially proposed as a main building block of memory unit due to its high tunneling magnetoresistance ratio\cite{ikeda_2008} and current-induced switching\cite{huai_2004}. Unfortunately, in general STT-based devices require high critical current densities via thin tunnel barrier for the magnetisation switching and suffer from low thermal stability. Perpendicularly magnetised MTJs using both interface perpendicular magnetic anisotropy (PMA) \cite{nakayama_2008,ikeda_2010,amiri_2011} and shape anisotropy\cite{watanabe_2018} has been proposed to enhance thermal stability and to reduce critical currents densities, however, MgO tunnel barrier can still degrade its own properties over the time due to high switching currents. To overcome this problem, the SOT effect, in which spin currents are generated in layers with high spin-orbit couplings, such as heavy metals (HM)\cite{liu_2012} and topological insulators (TI)\cite{DC_2018} has been studied. The switching of the magnetisation driven by SOT with PMA does not require such a high current density tunneling through a MgO barrier \cite{liu_2012}. Originally, an external magnetic field had to be applied to assist SOT switching in order to break the time-reversal symmetry. Although the need for external field in device applications would not be practical, therefore, numerous alternatives have been proposed, such as coupling to the antiferromagnets\cite{mendes_2014,ou_2016,oh_2016,fukami_2016,wu_2016,razavi_2017} or to another ferromagnet \cite{liu_2019,lazarski_2019}, use of interface torques\cite{baek_2018} and combination of STT and SOT effects\cite{wang_2018}. The use of trilayer structure with two ferromagnet with different magnetic anisotropies coupled by Ta layer forming so-called T-type structure has led to the observation of robust field-free switching\cite{kong_2019}. 

In this work we study the Co$_{20}$Fe$_{60}$B$_{20}$/Ta /Co$_{20}$Fe$_{60}$B$_{20}$ trilayer structure in order to determine the coupling and SOT-induced magnetisation switching. Unlike in Ref.~[\onlinecite{kong_2019}] a single Ta layer serves both as the coupling layer and the spin current source. By carefully selecting the CoFeB layer thicknesses, deposition condition and thermal treatment we are able to control the magnetic anisotropy. The interlayer exchange coupling (IEC) across a Ta spacer is investigated using the ferromagnetic resonance (FMR) measurements.\cite{yu_2014,you_2015} This coupling induces different magnetic orientation of two CoFeB layers, which results in a non-trivial dependence of the switching field and critical SOT-induced current dependence as a function of Ta thickness.
 
\section{Experiment}
All samples were magnetron sputtered on thermally oxidized Si wafer, with a  wedged shape Ta layer deposited using a moving shutter technique. The investigated multilayer structure is as following:
Ta(2)/Co$_{20}$Fe$_{60}$B$_{20}$(4)/ Ta($t_\mathrm{Ta}$) /Co$_{20}$Fe$_{60}$B$_{20}$(1.3)/MgO(3)/TaOx(2) (thickness in nm) with $t_\mathrm{Ta}$ varying from 0 to 8 nm.
The bottom (thicker) and top (thinner) Co$_{20}$Fe$_{60}$B$_{20}$ layers exhibit in-plane (IPMA) and perpendicular magnetic anisotropy (PMA), respectively. Additional multilayers with single ferromagnetic layer (FM) and variable thicknesses of Ta or CoFeB were deposited: Ta(0-10)/Co$_{20}$Fe$_{60}$B$_{20}$(5)/Ta(1) and Ta(5)/Co$_{20}$Fe$_{60}$B$_{20}$(0-10)/Ta(1) for resistivity analisys and PMA measurements using anomalous Hall effect (AHE). Weak effective PMA of the top CoFeB layer was observed after the deposition process for a limited thickness of CoFeB around 1 nm. After annealing at 300$^{\circ}$C, the range of CoFeB thicknesses for which effective PMA was found increased, which is confirmed by the evolution of the AHE signal in a half-stack with variable thickness of FM: Ta(5)/Co$_{20}$Fe$_{60}$B$_{20}$(0-2)/MgO(2)/Ta(1). The resistivity of CoFeB and Ta were determined in bilayers of variable thickness according to the model presented in Ref.~[\onlinecite{kawaguchi_2018}]. After the deposition process, all samples were characterised by X-ray diffraction (XRD) and X-ray reflectivity (XRD) measurements using X'Pert-MPD diffractometer with Cu-anode to assist in controlling  the thickness of particular layers in the system and to analyse their crystal orientation. \newline

For the thinnest wedge thicknesses of $t_\mathrm{Ta}$ < 2 nm, the grazing incident angle profile measurements show no peaks. For 2 < $t_\mathrm{Ta}$ < 6 nm, a broad peak corresponding to the amorphous Ta appears. Finally, for $t_\mathrm{Ta}$ > 6 nm the measurement results contain peaks that originate from a disoriented polycrystalline tetragonal $\beta$ phase.

\begin{figure}[H]
\centering
\includegraphics[width=\columnwidth]{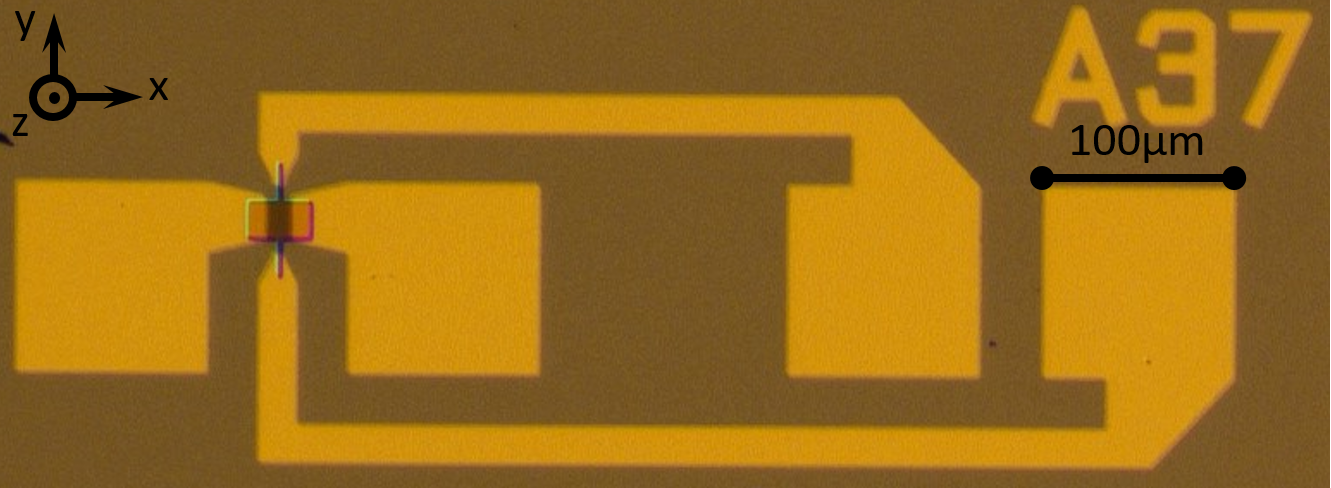}
\caption{Micrograph of the Hall-bar structure used for SOT current-induced magnetisation switching measurements.}
\label{fig:structures}
\end{figure}

For magneto-transport measurements, the Hall bars along with the rectangular stripes were fabricated using 385-nm projection lithography, ion-beam etching and lift-off process. The elements dimensions were: $10~\mu\mathrm{m}$ $\times$ $100~\mu\mathrm{m}$ for SOT dynamics measurements, $30~\mu\mathrm{m}$ $\times$ $10~\mu\mathrm{m}$ stripes for current-induced magnetisation switching and Hall bars $10~\mu\mathrm{m}$ $\times$ $80~\mu\mathrm{m}$ for AHE measurements. An example of the fabricated Hall bar with additional contact electrodes made of Al(20)/Au(30) is presented in Fig.\ref{fig:structures}. \newline

AHE was measured in a perpendicular magnetic field with constant current flowing through the long axis of the Hall bar. Spin-orbit torque ferromagnetic resonance (SOT-FMR) with RF signal of the constant frequency was applied and DC voltage was measured at the same electrodes with in-plane magnetic field swept along 45$^\circ$ axis with respect to the long axis of the stripe.

\section{Result and discussion}
	\subsection{Anomalous Hall effect}
\begin{figure}[H]
\centering
\includegraphics[width=\columnwidth]{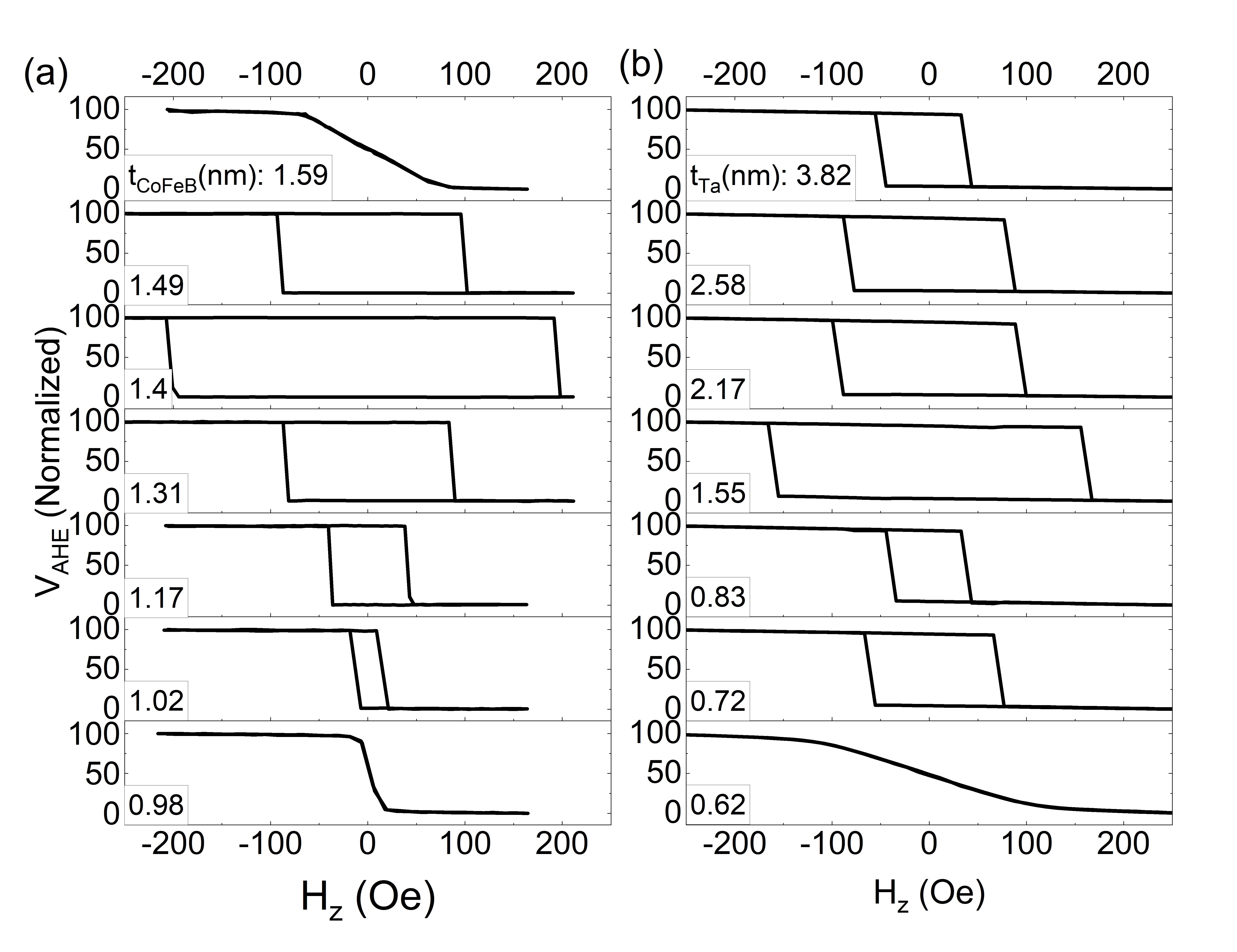}
\caption{AHE hysteresis loops measured in: (a) a bilayer with a single FM as a function of CoFeB thickness: Ta(5)/CoFeB(0-2)/MgO(2)/Ta(1), (b) trilayers with two FM layers separated by different Ta spacer thickness:  Ta(2)/CoFeB(5)/Ta(0-10) /CoFeB(1)/MgO(2)/Ta(1). Both structures were annealed at 300$^{\circ}$C. }
\label{fig:AHE_vol1}
\end{figure}

CoFeB and Ta layers resistivity were determined as described in Ref.~[\onlinecite{kawaguchi_2018}], and the results were calculated to be $\rho_{\text{Ta}}$ = 170$\mu\Omega$cm and $\rho_{\text{CoFeB}}$ = 160$\mu\Omega$cm. Magnetic anisotropy energy was determined from the AHE measurements. The maximum coercivity was measured for $t_\mathrm{CoFeB}$ = 1.3-1.5 nm. Thinner (thicker) CoFeB are characterised by perpendicular (in-plane) magnetic anisotropy what is shown in Fig.\ref{fig:AHE_vol1}(a).
Fig.\ref{fig:AHE_vol1}(b) presents the AHE loop evolution of CoFeB/Ta/CoFeB trilayers as a function of the Ta spacer thickness. 
Clearly, the magnetic behaviour of the top CoFeB layer is modulated by the spacer thickness. For $t_\mathrm{Ta}$ below 0.7 nm, hard axis AHE hysteresis loop corresponding to the ferromagnetic coupling to the bottom in-plane magnetised CoFeB layer is measured. For thicker Ta the coercive field of the top CoFeB is modulated, which is a signature of the antiferromagnetic coupling of different energy, as proposed in Ref.~[\onlinecite{parkin_1991}].

However, the switching process in micron-scale Hall bars is determined by the domain nucleation and propagation, therefore, precise determination of the coupling energy is challenging. For this purpose, we performed additional SOT-FMR measurements of the coupled CoFeB/Ta/CoFeB system. 


	\subsection{Interlayer exchange coupling}
\begin{figure}[H]
\centering
\includegraphics[width=\columnwidth]{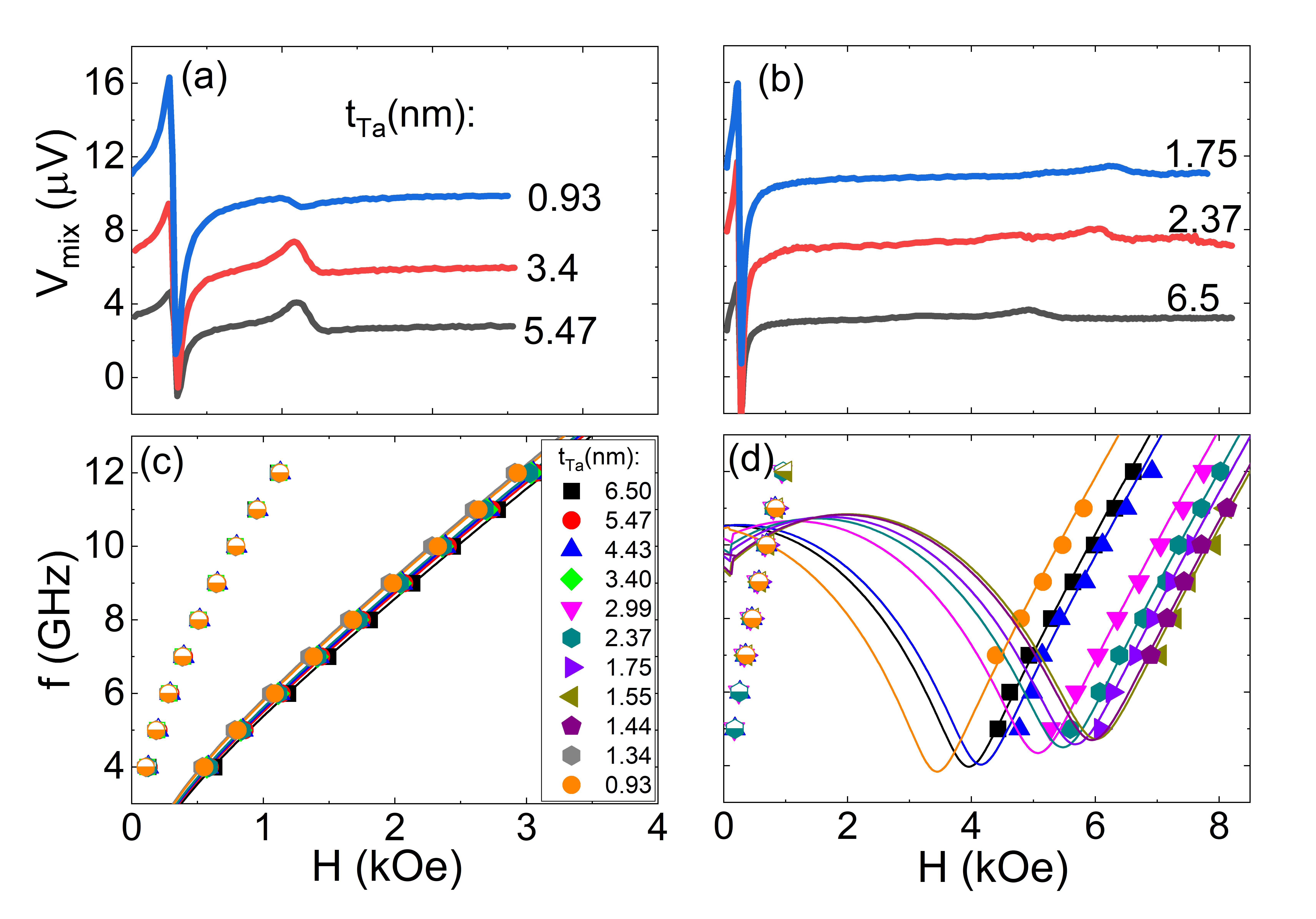}
\caption{(a) and (b) show exemplary SOT-FMR measurement in as-deposited and annealed states, respectively. The Kittel relation for both states - (c) and (d).}
\label{fig:VFMR}
\end{figure}

An example of the SOT-FMR measurement for a trilayer in as-deposited and annealed state is presented in Fig.\ref{fig:VFMR}(a-b).
Two peaks in the magnetic field-domain correspond to the resonance frequency of the in-plane magentised thick CoFeB (low field) and out-of-plane magnetised thin CoFeB (high field). Extracted signal, $V_\mathrm{mix}$, consists of two parts - symmetric and antisymmetric - which are modeled by Lorentz curves:
 \begin{equation}
V_{\mathrm{mix}} = V_\mathrm{S}\frac{\Delta H^2}{\Delta H^2+(H-H_0)^2} + V_\mathrm{A} \frac{\Delta H(H-H_{0})}{\Delta H^2+(H-H_{0})^2}\,
\label{eq:Vmix}
\end{equation}
where $V_\mathrm{S}$ ($V_\mathrm{A}$) is the magnitude of symmetrical (antisymmetrical) Lorentz curve, $\Delta H$ is the linewidth, and $H_\mathrm{0}$ is the resonance field. \newline
 
Fig.\ref{fig:VFMR}(c-d) presents the $f(H)$ dependences measured for different thickness of the Ta spacers. Strong dependence of the peak position on $t_\mathrm{Ta}$ is observed in the annealed sample, which indicates that the coupling is greatly enhanced after the thermal treatment\cite{liu_2016}.
The dependence was analysed using the macro-spin model, taking into account the following parameters: saturation magnetisation, $\mu_0M_\mathrm{S}$ = 1.5 T, anisotropy of bottom and top layer, $K_\mathrm{A1}$ = $2.8*10^3 J/m^3$ and $K_\mathrm{A2}$ = $1.12*10^6 J/m^3$ with the only variable being the coupling strength.
The magnetic anisotropy values of the bottom and top CoFeB layers correspond to the effective anisotropies of $K_\mathrm{eff_1}=9.04*10^5 J/m^3$ and $K_\mathrm{eff_2} = -2.13*10^5 J/m^3$ based on Eq.~\ref{eq:Keff}:

\begin{equation}
K_{eff}= \frac{\mu_0{M_s}^2}{2}-K_{A},
\label{eq:Keff}
\end{equation}
where $\mu_0$ is magnetic permeability.

\begin{figure}[H]
\centering
\includegraphics[width=\columnwidth]{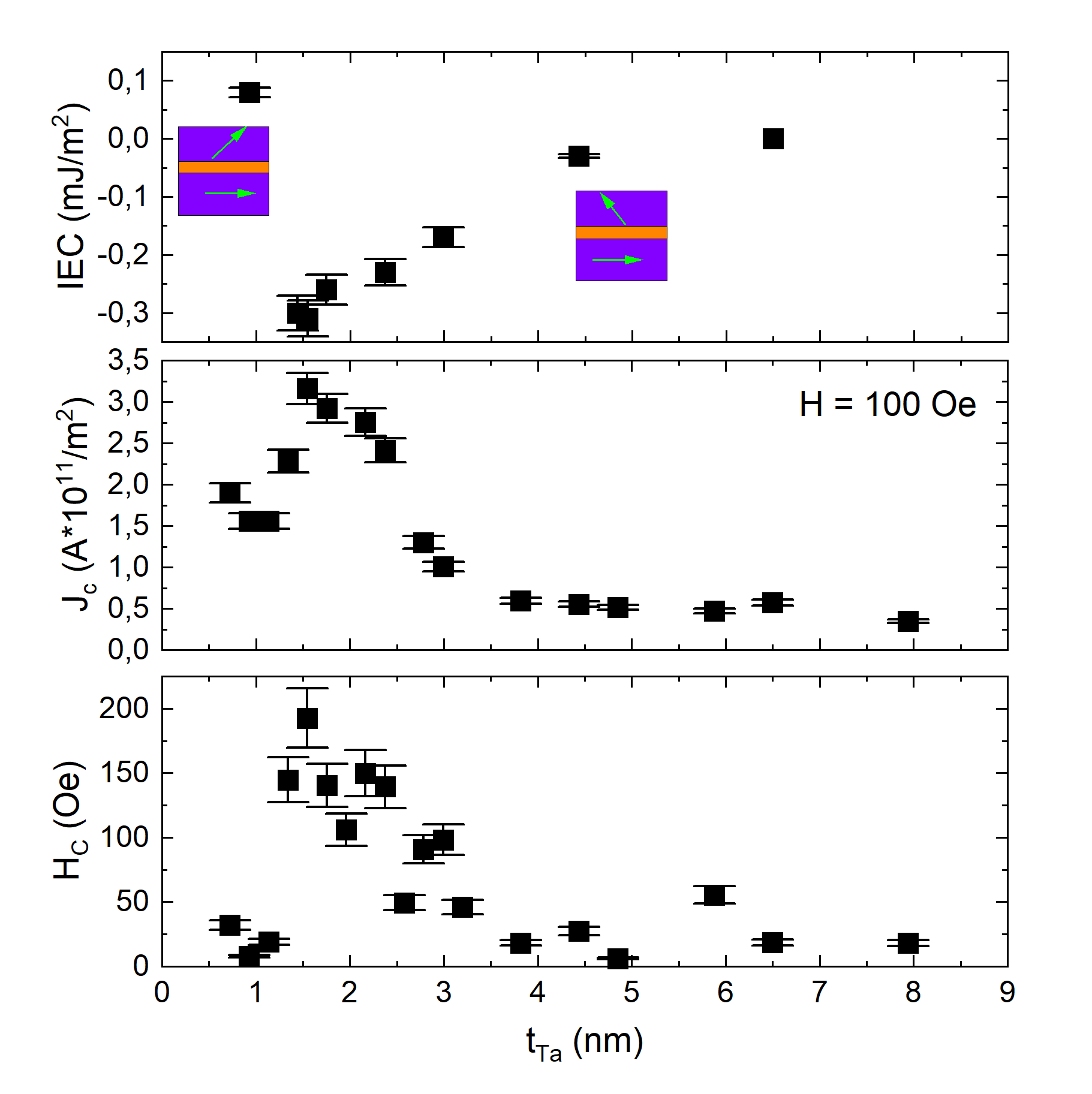}
\caption{(a) IEC energy, (b) critical current density ($J_\mathrm{C}$) and (c) coercivity field ($H_\mathrm{C}$) as a function of $t_\mathrm{Ta}$. }
\label{fig:IEC}
\end{figure}

The coupling energy determined from the macrospin model as a function of $t_\mathrm{Ta}$ is presented in Fig.\ref{fig:IEC}(a). Contrary to Ref.~[\onlinecite{cheng_2012}] in the mixed anisotropy system we were not able to observe the oscillating behaviour of the coupling energy, nevertheless the maximum antiferromagnetic coupling corresponds to similar $t_\mathrm{Ta}$ of around 1.5 nm.



	\subsection{SOT induced switching}
\begin{figure}[H]
\centering
\includegraphics[width=\columnwidth]{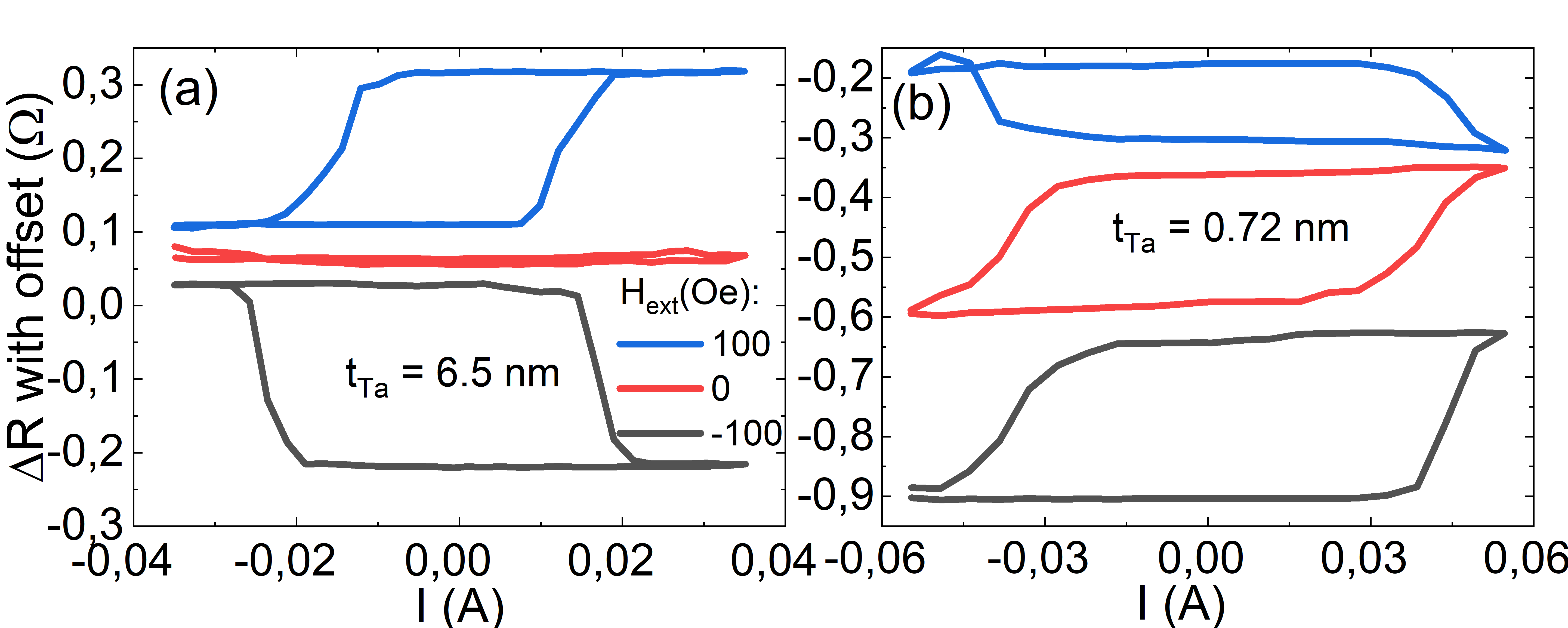}
\caption{(a) SOT-induced switching with antiferromagnetic coupling. (b) Field free switching achieved for the sample in ferromagnetic regime.}
\label{fig:SOT_switchning}
\end{figure}

After the analysis of the coupling energy, we turn to the SOT-induced switching experiments. Similary to the previous work~[\onlinecite{lazarski_2019}] we used HM both as a spacer and a source of the spin current. Unlike in the Co/Pt/Co case, where strong ferromagnetic coupling was observed in the entire Pt thickness range\cite{bonda_2020}, the CoFeB/Ta/CoFeB did not result in a field-free switching in the antiferromagnetic coupling region. The example of the switching process with the assistance of the external magnetic field is presented in Fig.\ref{fig:SOT_switchning}(a). The dependence of the switching current density on $t_\mathrm{Ta}$ in the external field of $H_\mathrm{x}$ = 100 Oe applied along the long axis of the Hall bar is presented in Fig.\ref{fig:IEC}(b). The behaviour closely resembles the dependence of the coercive field on the $t_\mathrm{Ta}$. 
\newline

The critical current density needed to switch the top perpendicular CoFeB magentisation in antiferromagnetic state increases along with an increase in coupling energy until the transition to the ferromagnetic coupling for $t_\mathrm{Ta}$ = 1.3 nm occurs. The structure switches deterministically only in the in-plane magnetic field ($H$ = +/- 100 Oe in the presented example). In the ferromagnetic coupling region, we observe current induced switching without any external magnetic field - Fig.\ref{fig:SOT_switchning}(b) - but not fully complete due to precisely selected ferromagnetic coupling  (similar to Ref.~[\onlinecite{lazarski_2019}]). A very strong ferromagnetic coupling appears in a narrow thickness of thin Ta layer. The switching process in this case may also be induced from the bottom Ta layer. Due to a limited range of Ta spacer thickness, for which the ferromagnetic coupling is observed, there is a need of additional spin-current source layer, in order to realize magnetic field-free switching\cite{kong_2019}. \newline


\section{Summary}
To sum up, the dynamics and the switching process for trilayer CoFeB/Ta/CoFeB structure with in-plane and perpendicular anisotropy of were investigated. The IEC was tuned by varying Ta layer thickness and changes from ferromagnetic ($t_\mathrm{Ta}$ < 1nm) to antiferromagnetic (1nm < $t_\mathrm{Ta}$ < 8nm ). In the antiferromagnetic regime, the small external magnetic field is needed to achieve the SOT-switching. On the contrary, in the ferromagnetic regime the field-free SOT-switching is gained but the range of change in resistance is slightly reduced. The implementation of two coupled FM layers with mixed anisotropies spaced by HM helps to understand new fundamental SOT mechanisms and could advance the development of field-free SOT-MRAM devices in the future.

\section*{Acknowledgments}
This work is supported by the National Science Centre, grant No. UMO-2015/17/D/ST3/00500, Poland. W.S. acknowledges grant No. LIDER/467/L-6/14/NCBR/2015 by the Polish National Centre for Research and Development. S. \L{}., K.G. and T. S. acknowledge National Science Centre grant Spinorbitronics UMO-2016/23/B/ST3/01430. Nanofabrication was performed at the Academic Centre for Materials and Nanotechnology of AGH University of Science and Technology. We thank Maciej Czapkiewicz for helpful
discussions on data analysis and Feliks Stobiecki for multilayers deposition.

\end{document}